\documentclass[aps,pra,superscriptaddress,reprint,floatfix]{revtex4-1}

\usepackage{amsmath}    
\usepackage{graphicx}   
\usepackage[dvipdfx,cmyk]{xcolor}     
\usepackage[colorlinks,linkcolor=red,citecolor=brown,urlcolor=blue]{hyperref}
\usepackage{dsfont,bm}
\usepackage{color}
\usepackage{pifont}
\usepackage{placeins}
\usepackage[T1]{fontenc} 
\definecolor{dgreen}{RGB}{0,120,0}

\usepackage{amsmath} 
\usepackage{amsthm} 
\usepackage{amssymb}	
\usepackage{graphicx} 
\usepackage{mathtools} 

\newcommand{\avg}[1]{\left< #1 \right>} 

\newcommand{\ket}[1]{\left| #1 \right>} 
\newcommand{\bra}[1]{\left< #1 \right|} 
\let\baraccent=\= 
\renewcommand{\=}[1]{\stackrel{#1}{=}} 

\theoremstyle{definition}

\theoremstyle{remark}

\graphicspath{{Figures/}}

\begin{document} 
\title{Deterministic delivery of remote entanglement on a quantum network}

\author{Peter~C.~Humphreys}
\author{Norbert~Kalb}
\affiliation{These authors contributed equally to this work.}
\affiliation{QuTech \& Kavli Institute of Nanoscience, Delft University of Technology, PO Box 5046, 2600 GA Delft, The Netherlands}
\author{Jaco~P.~J.~Morits}
\author{Raymond~N.~Schouten}
\author{Raymond~F.~L.~Vermeulen}
\affiliation{QuTech \& Kavli Institute of Nanoscience, Delft University of Technology, PO Box 5046, 2600 GA Delft, The Netherlands}
\author{Daniel~J.~Twitchen}
\author{Matthew~Markham}
\affiliation{Element Six Innovation, Fermi Avenue, Didcot, Oxfordshire OX11 0QE, U.K.}
\author{Ronald~Hanson}
\affiliation{QuTech \& Kavli Institute of Nanoscience, Delft University of Technology, PO Box 5046, 2600 GA Delft, The Netherlands}
\begin{abstract}
Large-scale quantum networks promise to enable secure communication, distributed quantum computing, enhanced sensing and fundamental tests of quantum mechanics through the distribution of entanglement across nodes~\cite{kimble_quantum_2008,broadbent_universal_2009,jiang_quantum_2009,ekert_ultimate_2014,gottesman_longer-baseline_2012,nickerson_freely_2014,komar_quantum_2014}. Moving beyond current two-node networks~\cite{Hucul2015,Hensen2015,Kalb2017,Reiserer2015,Hofmann72,northup2014quantum} requires the rate of entanglement generation between nodes to exceed their decoherence rates. Beyond this critical threshold, intrinsically probabilistic entangling protocols can be subsumed into a powerful building block that deterministically provides  remote entangled links at pre-specified times. Here we surpass this threshold using diamond spin qubit nodes separated by 2 metres. We realise a fully heralded single-photon entanglement protocol that achieves entangling rates up to 39 Hz, three orders of magnitude higher than previously demonstrated two-photon protocols on this platform~\cite{Pfaff2014}. At the same time, we suppress the decoherence rate of remote entangled states to 5 Hz by dynamical decoupling. By combining these results with efficient charge-state control and mitigation of spectral diffusion, we are able to deterministically deliver a fresh remote state with average entanglement fidelity exceeding 0.5 at every clock cycle of $\sim$100~ms without any pre- or post-selection. These results demonstrate a key building block for extended quantum networks and open the door to entanglement distribution across multiple remote nodes.
\end{abstract}
\maketitle 

The power of future quantum networks will derive from entanglement that is shared between the network nodes. Two critical parameters for the performance of such networks are the entanglement generation rate $r_\text{ent}$ between nodes and the entangled-state decoherence rate $r_\text{dec}$. Their ratio, that we term the quantum link efficiency $\eta_{\text{\,link}} = r_\text{ent} / r_\text{dec}$~\cite{Monroe2014,Hucul2015}, quantifies how effectively entangled states can be preserved over the timescales necessary to generate them. Alternatively, from a complementary perspective, the link efficiency determines the average number of entangled states that can be created within one entangled state lifetime. Achieving a link efficiency of unity therefore represents a critical threshold beyond which entanglement can be generated faster than it is lost. Exceeding this threshold is central to allowing multiple entangled links to be created and maintained simultaneously, as required for the distribution of many-body quantum states across a network~\cite{Monroe2014,nickerson_freely_2014}.

\begin{figure*}
\begin{center}
\includegraphics[width=14.0cm]{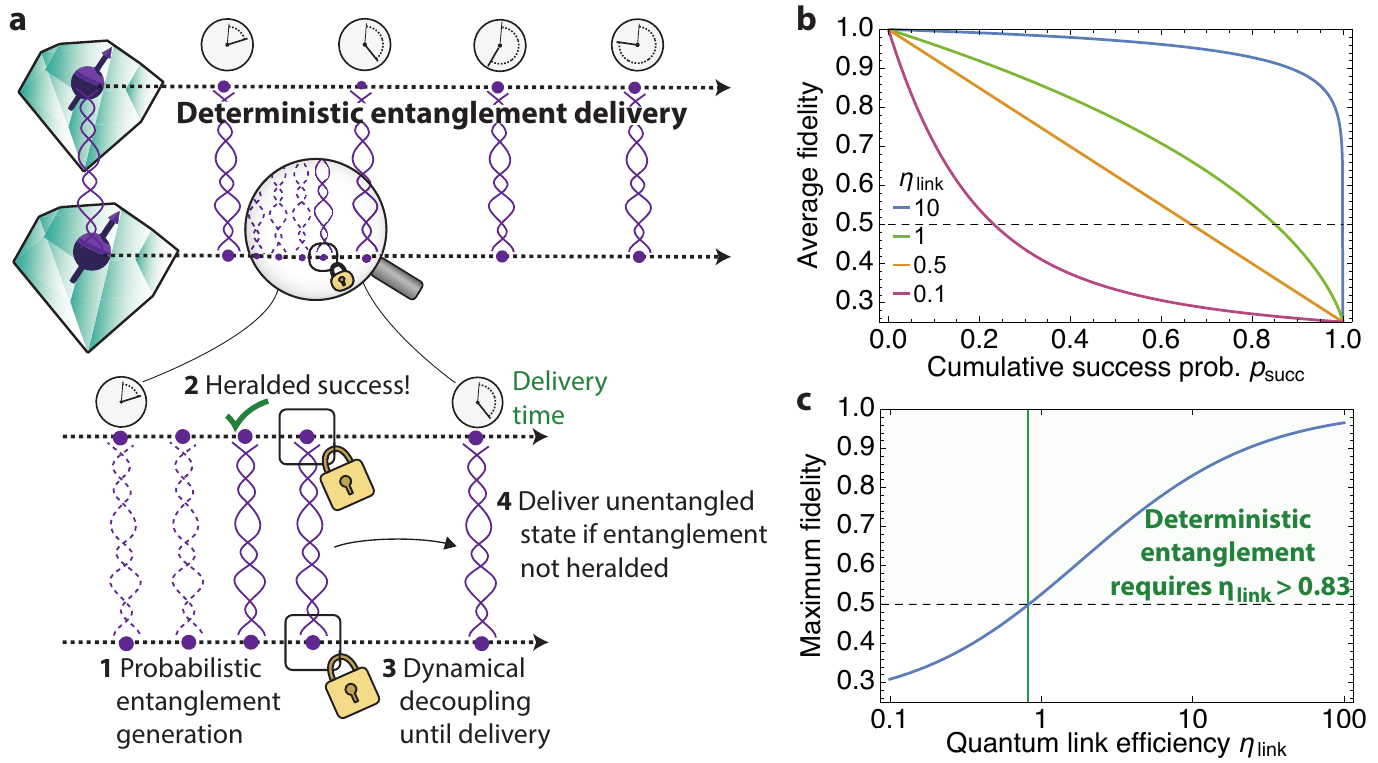}
\caption{\textbf{Deterministic remote entanglement delivery.} \textbf{a}, Deterministic entanglement delivery guarantees the output of states with average entanglement fidelity exceeding 0.5 at pre-specified times. In our protocol, underlying this deterministic delivery is a probabilistic but heralded entanglement process. Repeated entangling attempts are made and then, upon heralded success, the entangled state is protected from decoherence until the specified delivery time. If no attempt at entanglement generation succeeds within one cycle, an unentangled state must be delivered. \textbf{b}, For the underlying entanglement generation and state preservation protocol (steps 1, 2 \& 3 in (a)), the effectiveness of the trade-off between the average fidelity of the delivered entangled state and the success probability is determined by the quantum link efficiency $\eta_{\text{\,link}}$. \textbf{c}, Plotted is the maximum fidelity of deterministically delivered states as a function of $\eta_{\text{\,link}}$. A critical threshold $\eta_{\text{\,link}} \gtrsim 0.83$ must be surpassed in order to render the underlying probabilistic process deterministic and deliver an on-average entangled state at every cycle.}
\label{fig:QuantumCapacity}
\end{center}
\end{figure*}

Consider an elementary entanglement delivery protocol that delivers states at pre-determined times. This can be achieved by making multiple attempts to generate entanglement, and then protecting successfully generated entangled states from decoherence until the required delivery time (Fig.~\ref{fig:QuantumCapacity}a, steps 1, 2 \& 3). If we try to generate entanglement for a period $t_\text{ent}$, the cumulative probability of success will be $p_\text{succ} = 1-e^{- r_\text{ent} t_\text{ent}}$. For a given $p_\text{succ}$, the average fidelity $F_\text{succ}$ of the successfully generated states is solely determined by the quantum link efficiency $\eta_{\text{\,link}}$ (Supplementary Information (SI) section~I). We plot $F_\text{succ}$ versus $p_\text{succ}$ for several values of $\eta_{\text{\,link}}$ in Fig~\ref{fig:QuantumCapacity}b. 

This protocol allows entangled states to be delivered at specified times, but with a finite probability of success. By delivering an unentangled state (state fidelity $F_\text{unent} \leq \frac{1}{2}$) in cycles in which all entanglement generation attempts failed, the protocol can be cast into a fully deterministic black-box (Fig~\ref{fig:QuantumCapacity}a, step 4). The states output from such a black-box will have a fidelity with a Bell state of
\begin{equation}
F_\text{det} = p_\text{succ} F_\text{succ} + (1 - p_\text{succ}) F_\text{unent}.
\end{equation}
The maximum achievable fidelity $F_\text{det}^{\text{max}}$ of this deterministic state delivery protocol, found by optimising $p_\text{succ}$, is also only determined by the quantum link efficiency~$\eta_{\text{\,link}}$. For $F_\text{unent} = \frac{1}{4}$ (fully mixed state), we find (see Fig~\ref{fig:QuantumCapacity}c):
\begin{equation}
F_\text{det}^{\text{max}} = \frac{1}{4} \left(1+3 {\eta_\text{\,link}}^{\frac{1}{1-\eta_{\text{\,link}}}}\right).
\end{equation}
Beyond the threshold $\eta_{\text{\,link}} \gtrsim 0.83$, there exists a combination of $p_\text{succ}$ and $F_\text{succ}$ high enough to compensate for cycles in which entanglement is not heralded, allowing for the deterministic delivery of states that are on-average entangled ($F_\text{det}^{\text{max}}\geq\frac{1}{2}$). Demonstrating deterministic entanglement delivery therefore presents a critical benchmark of a network's performance, certifying that the network quantum link efficiency is of order unity or higher. Furthermore, the ability to specify in advance the time at which entangled states are delivered may assist in designing multi-step quantum information tasks such as entanglement routing~\cite{2017arXiv170807142P,2016arXiv161005238S}.

To date, this threshold has remained out of reach for solid-state quantum networks.  Quantum dots have demonstrated kHz entanglement rates $r_\text{ent}$, but tens of MHz decoherence rates $r_\text{dec}$ limit their achieved quantum link efficiencies to ${\eta_{\text{\,link}}\sim10^{-4}}$~\cite{Stockill2017,Delteil2016}. Nitrogen vacancy (NV) centres, point-defects in diamond with a long-lived electron spin and bright optical transitions, have demonstrated entanglement rates $r_\text{ent}$ of tens of mHz~\cite{Kalb2017,Pfaff2014} and, in separate experiments, decoherence rates $r_\text{dec}$ of order 1 Hz~\cite{bar2013solid}, which would together give $\eta_{\text{\,link}} \sim10^{-2}$.

\begin{figure*}
\begin{center}
\makebox[\textwidth][c]{\includegraphics[width=18cm]{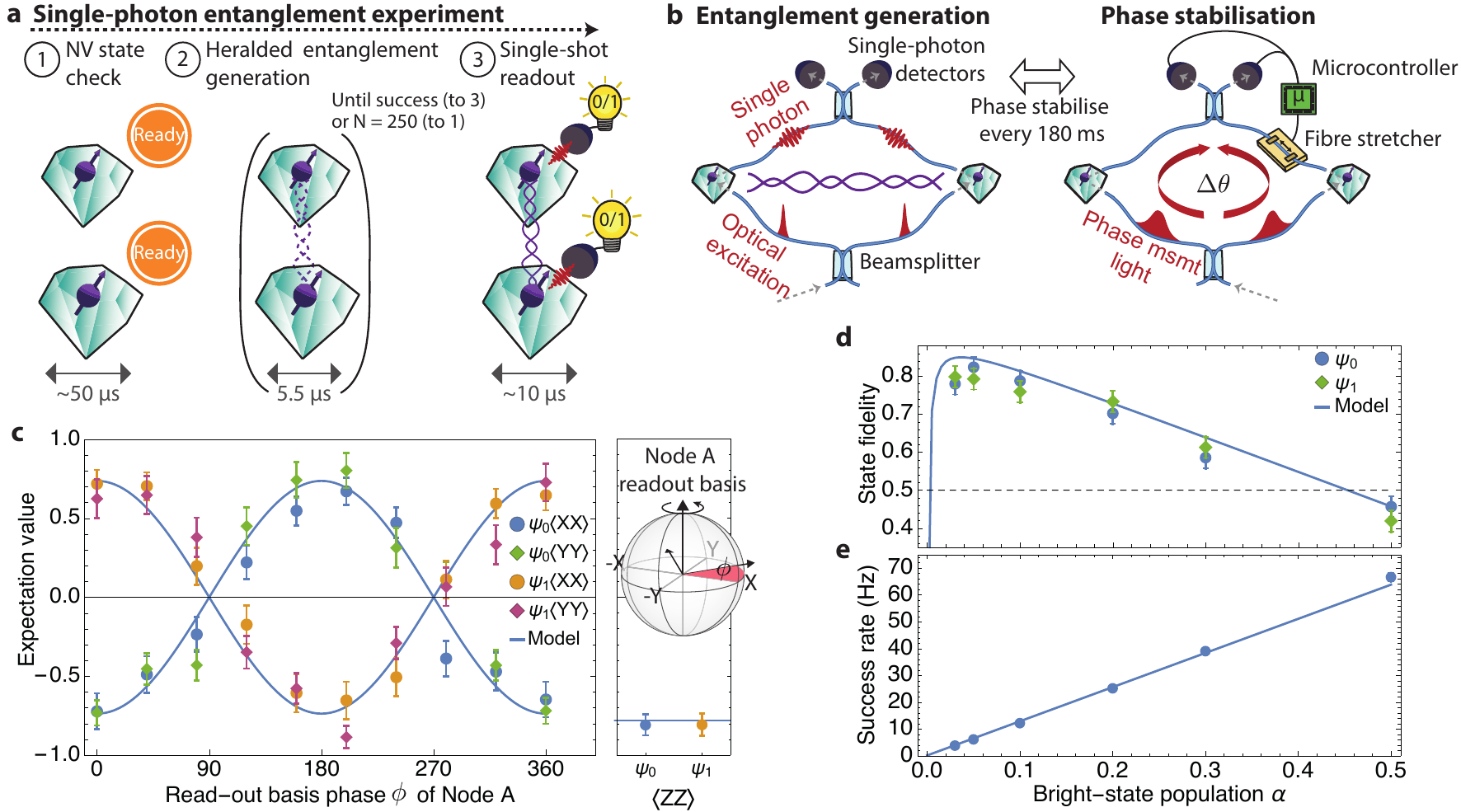}}%
\caption{\textbf{Benchmarking single-photon entanglement generation.} \textbf{a}, 
The single-photon entanglement experiment consists of several steps: \ding{172} Before entanglement generation, the NV centre must be in the correct charge state (NV$^-$) and resonant with the excitation laser. An NV state check is employed to ensure this (SI section II), and is repeated until the check passes. \ding{173} Single-photon entanglement generation is attempted until success is heralded, in which case we continue to readout. If 250 attempts have been made without success, we revert to step 1. \ding{174} Upon heralded success, the spin states are read out in a chosen basis by using microwaves to rotate the state followed by single-shot readout. \textbf{b}, The optical phase difference $\Delta \theta$ acquired in an interferometer formed by the two nodes must be known. For the data reported in this figure, we stabilise the phase difference every 180 ms. \textbf{c}, For $\alpha$ of 0.1, we plot the measured $\avg{XX}$ and $\avg{YY}$ correlations (left panel) for $\psi_{0/1}$ (where 0/1 denotes the heralding detector) as the phase of the microwave pulse before readout is swept at node A. This changes the readout basis of this node to $\cos{(\phi)} X + \sin{(\phi)} Y$. Right panel shows the measured $\avg{ZZ}$ correlations.
\textbf{d}, Fidelity of the heralded states with a Bell state and \textbf{e}, entanglement generation success rate, for different values of $\alpha$. The success rate is calculated by dividing the entangling attempt duration (5.5 $\mu s$) by the probability of successfully heralding entanglement. Solid lines in each plot give the predictions of our model solely based on independently determined parameters (SI section~III).}
\label{fig:SingleClickEntResults}
\end{center}
\end{figure*}

Here we achieve $\eta_{\text{\,link}}$ well in excess of unity by realising an alternative entanglement protocol for NV centres in which we directly use the state heralded by the detection of a single photon (Fig.~\ref{fig:SingleClickEntResults})~\cite{Cabrillo_1999,Minar2008}. The rate for such single-photon protocols scales linearly with losses, which, in comparison with previously used two-photon-mediated protocols~\cite{Pfaff2014,Hensen2015},  provides a dramatic advantage in typical remote entanglement settings. Recent experiments have highlighted the potential of such single-photon protocols by generating local entanglement~\cite{Casabone2013,Sipahigil2016}, and remote entanglement in post-selection~\cite{Stockill2017,Delteil2016}. By realising a single-photon protocol in a fully heralded fashion and protecting entanglement through dynamical decoupling, we achieve the deterministic delivery of remote entangled states on a $\sim$10~Hz clock.

Our experiment employs NV centres residing in independently operated cryostat setups separated by 2 metres (SI section~II). We use qubits formed by two of the NV centre ground-state spin sub-levels ($\ket{\uparrow} \equiv \ket{m_s = 0}, \ket{\downarrow} \equiv \ket{m_s=-1}$). Single-photon entanglement generation (Fig.~\ref{fig:SingleClickEntResults}a) proceeds by first initialising each node in $\ket{\uparrow}$ by optical pumping~\cite{Robledo2011}, followed by a coherent rotation using a microwave pulse~\cite{Fuchs1520} to create the state 
\begin{equation}
\ket{NV} = \sqrt{\alpha} \ket{\uparrow} + \sqrt{1-\alpha} \ket{\downarrow}.
\end{equation}
We then apply resonant laser light to selectively excite the `bright' state $\ket{\uparrow}$ to an excited state, which rapidly decays radiatively back to the ground state by emitting a single photon. This entangles the spin state of the NV with the presence $\ket{1}$ or absence $\ket{0}$ of a photon in the emitted optical mode:
\begin{equation}
\ket{NV, \text{optical mode}} = \sqrt{\alpha} \ket{\uparrow} \ket{1} + \sqrt{1-\alpha} \ket{\downarrow} \ket{0}.
\end{equation}

Emitted photons are transmitted to a central station at which a beamsplitter is used to remove their which-path information. Successful detection of a photon at this station indicates that at least one of the NVs is in the bright state $\ket{\uparrow}$ and therefore heralds the creation of a spin-spin entangled state. However, given the detection of one photon, the conditional probability that the other NV is also in the state $\ket{\uparrow}$, but the photon it emitted was lost, is given by $p =\alpha$ (in the limit that the photon detection efficiency $p_\text{det} \ll 1$). This degrades the heralded state from a maximally-entangled Bell state $\ket{\psi} = \frac{1}{\sqrt{2}}  (\ket{\uparrow \downarrow} + \ket{\downarrow \uparrow})$ to
\begin{equation}
\rho_{NV,NV} = (1-\alpha) \ket{\psi}\!\bra{\psi} + \alpha \ket{\uparrow \uparrow}\!\bra{\uparrow \uparrow}.\label{eqn:mixed}
\end{equation}

The probability of successfully heralding entanglement is given by $ 2 \, p_\text{det} \alpha$. The state fidelity $F = 1-\alpha$ can therefore be directly traded off against the entanglement rate. The corresponding success probability of a two-photon protocol is given by $\frac{1}{2} p_\text{det}^2$; for a given acceptable infidelity $\alpha$, single-photon protocols will thus provide a rate increase of $4 \, \alpha / p_\text{det}$. For example, for our system's ${p_\text{det}\sim4\times10^{-4}}$, if a 10\% infidelity is acceptable, the rate can be increased by three orders of magnitude over two-photon protocols. 

The primary challenge in implementing single-photon entanglement is that the resulting entangled state depends on the optical phase acquired by the laser pulses used to create spin-photon entanglement at each node, as well as the phase acquired by the emitted single photons as they propagate (Fig~\ref{fig:SingleClickEntResults}b). The experimental setup therefore acts as an interferometer from the point at which the optical pulses are split to the point at which the emitted optical modes interfere. For a total optical phase difference of $\Delta \theta$, the entangled state created is given by 
\begin{equation}
\ket{\psi_{0/1}(\Delta \theta)} = \ket{\uparrow \downarrow} \pm e^{\mathrm{i} \Delta \theta} \ket{\downarrow \uparrow},
\end{equation}
where $0/1$ (with corresponding $\pm$ phase factor) denotes which detector at the central station detected the incident photon. This optical phase difference must be known in order to ensure that entangled states are available for further use.

We overcome this entangled-state phase sensitivity by interleaving periods of optical-phase stabilisation with our entanglement generation. During phase stabilisation we input bright laser light at the same frequency as the NV excitation light and detect the light reflected from the diamond substrate using the same detectors that are used to herald entanglement. The measured optical phase, estimated from the detected counts, is used to adjust the phase back to our desired value using a piezoelectric fibre stretcher. We achieve an average steady-state phase stability of $14.3(3)^\circ$, limited by the mechanical oscillations of the optical elements in our experimental setup (SI section~V).

To demonstrate the controlled generation of entangled states, we run the single-photon entangling protocol with a bright-state population of $\alpha = 0.1$. After entanglement is heralded, we apply basis rotations and single-shot state readout~\cite{Robledo2011} at each node to measure $\avg{\sigma_i^A \sigma_j^B}$ correlations between the nodes, where the standard Pauli matrices will be referred to here in the shorthand $\sigma_{X},\sigma_{Y},\sigma_{Z} = X,Y,Z$. We observe strong correlations both for $\avg{XX}$ and $\avg{YY}$, and, when sweeping the readout basis for node~A, oscillations of these coherences as expected from the desired entangled state (Fig.~\ref{fig:SingleClickEntResults}c, left panel). In combination with the measured $\avg{ZZ}$ correlations (Fig.~\ref{fig:SingleClickEntResults}c, right panel), this unambiguously proves the establishment of entanglement between our nodes.

We explore the tradeoff between the entangled state fidelity and the entanglement rate by measuring $\avg{XX}$, $\avg{YY}$ and $\avg{ZZ}$ correlations for a range of different initial bright-state populations $\alpha$. Using these correlations, we calculate the fidelity of the heralded state to the desired maximally entangled Bell state for each value of $\alpha$ (Fig.~\ref{fig:SingleClickEntResults}d), along with the measured success rate (Fig.~\ref{fig:SingleClickEntResults}e). As predicted, the fidelity increases with decreasing $\alpha$ as the weight of the unentangled state $\ket{\uparrow \uparrow}\!\bra{\uparrow \uparrow}$ diminishes (Eqn.~\ref{eqn:mixed}). For small $\alpha$, the fidelity saturates because the detector dark-count rate becomes comparable to the detection rate. 

Choosing $\alpha$ to maximise fidelity, we find that our protocol allows us to generate entanglement with a fidelity of $0.81(2)$ at a rate of $r_\text{ent} = 6$ Hz (for $\alpha = 0.05$). Alternatively, by trading the entanglement fidelity for rate, we can generate entanglement at $r_\text{ent} = 39$ Hz with an associated fidelity of $0.60(2)$ ($\alpha = 0.3$). This represents a two orders of magnitude increase in the entangling rate over all previous NV experiments~\cite{Kalb2017} and a three orders of magnitude increase in rates over two-photon protocols under the same conditions~\cite{Pfaff2014}. 

\begin{figure}
\begin{center}
\includegraphics[width=8.8cm]{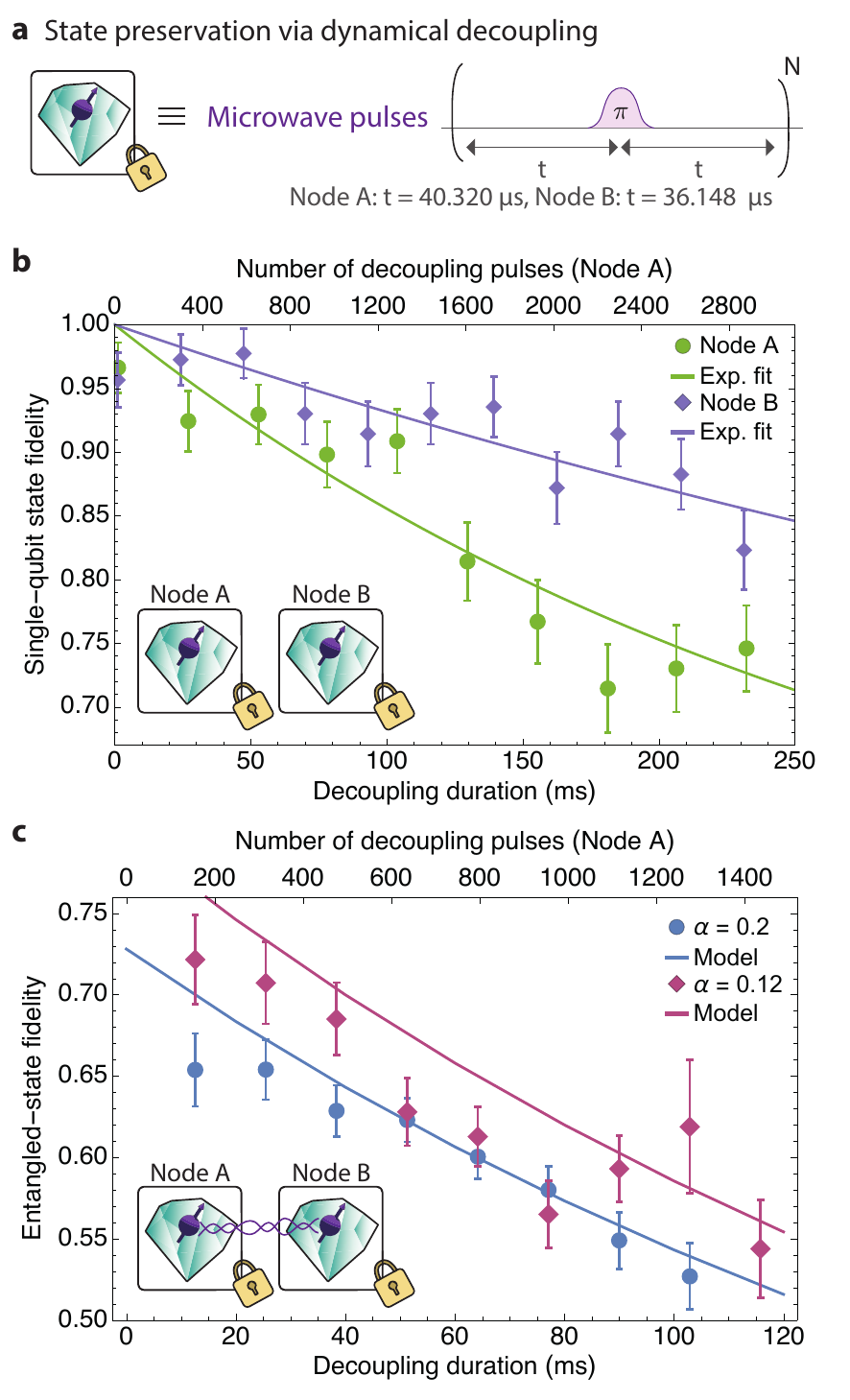}
\caption{\textbf{Coherence protection of remote entangled states.} \textbf{a}, Dynamical decoupling protects the state of the NV spins from quasi-static environmental noise.  Applying $N$ pulses allows us to dynamically decouple the NV state for a time $2 N t$. \textbf{b}, Fidelity with the initial state for dynamical decoupling of the single-qubit state $\ket{\uparrow} + \ket{\downarrow}$ at our two NV nodes. Solid lines show exponential fits with coherence times of $290(20)$ ms and $680(70)$ ms for nodes A and B respectively. \textbf{c}, Dynamical decoupling of entangled states created using the single-photon entanglement protocol for bright-state populations $\alpha =0.12$ and $\alpha =0.2$. Solid lines show the predictions of our model based on the coherence times measured in (b), from which the effective entangled state coherence time is expected to be $\tau = 200(10)$~ms.}
\label{fig:SuccRateAndDD}
\end{center}
\end{figure}

Compared to the maximum theoretical fidelity for $\alpha = 0.05$ of 0.95, the states we generate have a 3\% reduction in fidelity due to residual photon distinguishability, 4\% from double excitation, 3\% from detector dark counts, and 2\% from optical-phase uncertainty (SI sections~V,~VI~\&~VII). 

In order to reach a sufficient link efficiency $\eta_{\text{\,link}}$ to allow for deterministic entanglement delivery, the single-photon protocol must be combined with robust protection of the generated remote entangled states. To achieve this, we carefully shielded our NVs from external noise sources including residual laser light and microwave amplifier noise, leaving as the dominant noise the slowly-fluctuating magnetic field induced by the surrounding nuclear spin bath. 

We mitigate this quasi-static noise by implementing dynamical decoupling with XY8 pulse sequences (Fig.~\ref{fig:SuccRateAndDD}a, SI section~VIII). The fixed delay between microwave pulses in these sequences is optimised for each node~\cite{2017abobeih}. Varying the number of decoupling pulses allows us to protect the spins for different durations. This dynamical decoupling extends the coherence time of Node A and B from a $T_2^*$ of $\sim5 \, \mu$s to $290(20)$~ms and $680(70)$~ms respectively, as shown in Fig.~\ref{fig:SuccRateAndDD}b. The difference in coherence times for the two nodes is attributed to differing nuclear spin environments and microwave pulse fidelities.

\begin{figure*}[p!]
\begin{center}
\includegraphics[width=14.0cm]{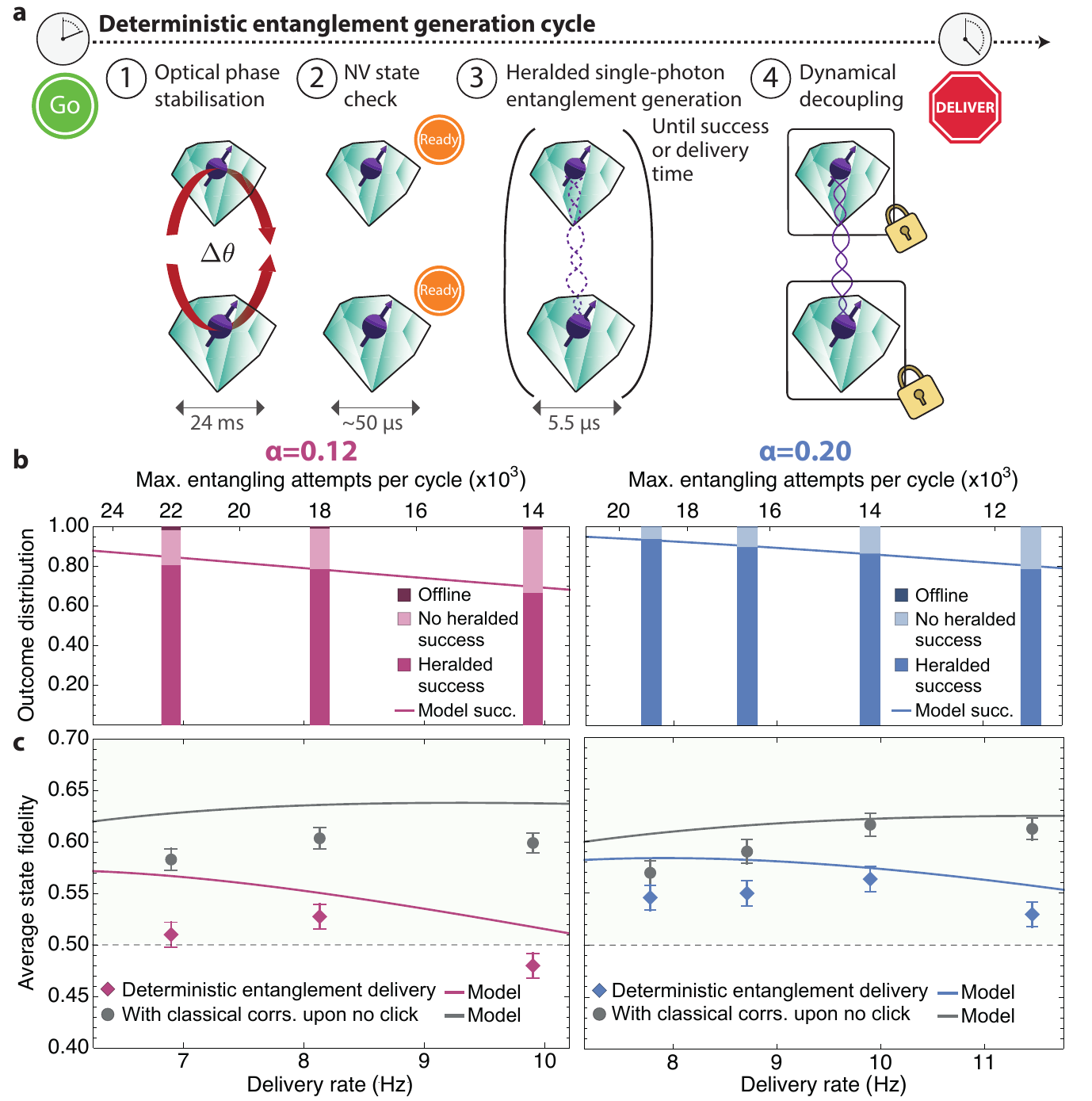}
\caption{\textbf{Deterministic entanglement delivery.}  \textbf{a}, One cycle of deterministic entanglement delivery combines the following steps: \ding{172} Optical phase stabilisation. \ding{173} NV state checks, repeated until a threshold number of photons are detected at each node. \ding{174} Attempts at probabilistic entanglement generation (Fig.~\ref{fig:SingleClickEntResults}). \ding{175} Upon heralded entanglement success, the state is protected by dynamical decoupling until the delivery time. \textbf{b}, Distribution of deterministic entanglement delivery outcomes for $\alpha$ = 0.12 \& 0.2 and different delivery rates. Shown is the fraction of cycles in which a herald photon is detected (heralded success), in which no herald is detected (no heralded success), and in which the NV state checks for at least one of the NV centres fail repeatedly for the whole cycle (offline). Note that the offline fraction is often too small to be visible in the plot. The line gives the success rate predicted by our model. \textbf{c}, Average fidelity of deterministically-delivered entangled states for $\alpha$ = 0.12 \& 0.2 and different delivery rates (diamonds). Also plotted is the average fidelity if classically-correlated states were delivered for cycles in which no success event is heralded (circles). The associated lines plot the corresponding predictions of our model based on independently measured parameters (SI section~III).}
\label{fig:DeterministicEntGenerationExpm}
\end{center}
\end{figure*}

To investigate the preservation of remote entangled states, we incorporate dynamical decoupling for varying time durations after successful single-photon entanglement generation (Fig.~\ref{fig:SuccRateAndDD}c). We find an entangled state coherence time of $200(10)$ ms (decoherence rate $r_\text{dec}$ of 5.0(3) Hz). The observed entangled-state fidelities closely match the predictions of our model, which is solely based on independently determined parameters (SI section~III). In particular, the decoherence of the remote entangled state is fully explained by the combination of the individual decoherence rates of the individual nodes. 

The combination of dynamical decoupling with the single-photon entanglement protocol achieves a quantum link efficiency of ${\eta_{\text{\,link}}\sim8}$ (comparable to the published state-of-the-art in ion traps, ${\eta_{\text{\,link}}\sim5}$~\cite{Hucul2015}), pushing the NV-based platform well beyond the critical threshold of $\eta_{\text{\,link}} \gtrsim 0.83$.

We capitalise on these innovations to design a deterministic entanglement delivery protocol that guarantees the delivery of entangled states at specified intervals, without any post-selection of results or pre-selection based on the nodes being in appropriate conditions (Fig.~\ref{fig:DeterministicEntGenerationExpm}a). Phase stabilisation occurs at the start of each cycle, after which there is a preset period before an entangled state must be delivered. This window must therefore include all NV state checks (necessary to mitigate spectral diffusion via feedback control and verify the charge-state and resonance conditions~\cite{Hensen2015}), entanglement generation attempts and dynamical decoupling necessary to deliver an entangled state (further details are given in SI section~II).

We run our deterministic entanglement delivery protocol at two values of $\alpha$ (0.2 \& 0.12) and for delivery rates ranging from 7-12 Hz. 
We divide the experiment into runs of 1500 cycles (i.e. 1500 deterministic state deliveries), for a total data set of 42000 cycles. 

We first confirm that heralded entanglement occurs with the expected probabilities (Fig.~\ref{fig:DeterministicEntGenerationExpm}a) by determining the fraction of cycles in which entanglement is heralded, in which no entangling attempts succeed, and in which entanglement attempts do not occur at all as the NV state check never succeeds.  In order to establish reliable and useful quantum networks, it is important that entangled states can be delivered with high confidence over long periods. The nodes must therefore not be offline due, for example, to uncompensated drifts in the resonant frequencies of the optical transitions. We therefore do not stop the experiment from running once it starts and include any such offline cycles in our datasets. Their negligible contribution (0.8\% of cycles) confirms the high robustness of our experimental platform and the effectiveness of our NV frequency and charge-state control (SI sections~II~\&~IV). 

For each value of $\alpha$ and for each pre-set delivery interval, we determine the average fidelity of the deterministically delivered states by measuring their $\avg{XX}$, $\avg{YY}$ and $\avg{ZZ}$ correlations (Fig.~\ref{fig:DeterministicEntGenerationExpm}b). We find that for $\alpha = 0.2$ and a rate of 9.9 Hz, we are able to create states with a fidelity of $0.56(1)$, proving successful deterministic entanglement delivery.

During cycles in which entanglement is not successfully heralded, the spin states are nonetheless delivered and readout. In these case, we deliver the state that the NVs are left in after a failed entanglement attempt, which has a low fidelity with the desired Bell state (e.g. $F_\text{unent} = 0.04$ for $\alpha = 0.2$). While this stringent test highlights the robust nature of our protocol, we could instead deliver a mixed state ($F_\text{unent}=\frac{1}{4}$) or a classically-correlated state ($F_\text{unent} = \frac{1}{2}$) when a successful event is not heralded.  The resulting fidelities for our experimental data if classically-correlated states were delivered are also plotted in Fig.~\ref{fig:DeterministicEntGenerationExpm}b (grey circles). In this case we would be able to deliver entangled states deterministically with fidelities of $0.62(1)$ at a rate of 9.9 Hz.

The deterministic entanglement delivery between remote NV centres demonstrated here is enabled by a quantum link efficiency exceeding unity.  Straightforward modifications to our experiment are expected to further increase our quantum link efficiency. Refinements to the classical experimental control will allow us to reduce the entanglement attempt duration from 5.5 $\mu$s to below 2 $\mu$s, which would more than double the entangling rate. Furthermore, the entangled state coherence time could be significantly improved by exploiting long-lived nuclear spin quantum memories~\cite{Maurer2012,Kalb2017,yang2016high}. We anticipate that this will allow for link efficiencies in excess of 100 in the near term. Further improvements to the photon detection efficiency (including enhancement of zero-phonon line emission)~\cite{Riedel_2017,2017arXiv171101704W} would lead to an additional increase of at least an order of magnitude.

In combination with recent progress on robust storage of quantum states during remote entangling operations~\cite{Reiserer2016,Kalb2017}, the techniques reported here reveal a direct path to the creation of many-body quantum states distributed over multiple quantum network nodes. Moreover, given the demonstrated potential for phase stabilisation in optical fibre over tens of kilometre distances~\cite{Minar2008}, our results open up the prospect of entanglement-based quantum networks at metropolitan scales.

\section*{Acknowledgements}
We thank Suzanne van Dam, Mohamed Abobeih, Tim Taminiau, Filip Rozp\k{e}dek, Kenneth Goodenough and Stephanie Wehner for helpful discussions. We acknowledge support from the Netherlands Organisation for Scientific Research (NWO) through a VICI grant and the European Research Council through a Starting Grant and a Synergy Grant.

Correspondence and requests for materials should be addressed to R.H. (r.hanson@tudelft.nl).

\bibliographystyle{naturemag}
\bibliography{bib_library}
\end{document}